\documentclass[aip,pof,reprint]{revtex4-1}
\newlength{\figwidth}
\usepackage{graphicx}
\def\apj{Astrophysical Journal}
\def\mnras{Monthly Notices of the Royal Astronomical Society}

\draft 

\begin{document}


\title{The late time structure of high density contrast, single mode
  Richtmyer-Meshkov flow} 



\author{R. J. R. Williams}
\affiliation{AWE Aldermaston, Reading, Berkshire, RG7 4PR, United Kingdom}


\date{\today}

\begin{abstract}
  We study the late time flow structure of Richtmyer-Meshkov
  instability.  Recent numerical work\cite{cherne2015shock} has
  suggested a self-similar collapse of the development of this
  instability at late times, independent of the initial surface
  profile.  Using the form of collapse suggested, we derive an
  analytic expression for the mass-velocity relation in the spikes,
  and a global theory for the late time flow structure.  We compare
  these results with fluid dynamical simulation.
\end{abstract}

\pacs{}

\maketitle 

\section{Introduction}

Richtmyer-Meshkov instability is the process by which perturbations
grow when a surface is subject to an impulsive acceleration.  The
growth of perturbations at shocked interfaces is relevant to a wide
range of contexts, from inertial confinement fusion\cite{lindl} to the
growth of mixing in
supernovae\cite{1955ApJ...122....1L,remington2000review}.  In this
paper, we discuss the Atwood number unity limit of this instability,
where one material is far more dense than the other.  We consider the
growth of perturbations from a surface with periodic perturbations, as
might result from machining, and discuss the late time asymptotic
structure which results in this case.  We assume that after the
material has shocked, it can be treated as an ideal fluid.  In
practical applications, the periodicity of the resulting flow would be
expected to break down at late time due to the effects of small
surface irregularities, and real fluid properties such as viscosity,
strength and surface tension would be expected to become increasingly
important in the flow \cite{sohn2009effects}.  Nevertheless, the
asymptotic structure of the flow for an ideal fluid should be a useful
indication of the behaviour one might expect for a real material at an
intermediate time.

We first review experimental and theoretical work on the growth of
single-mode Rayleigh-Taylor (RT) and Richtmyer-Meshkov (RM)
instability which provide a context for the present work.  Using a
scaling suggested by recent work\cite{cherne2015shock}, we derive
self-similar equations for the late time asymptotic structure of
Atwood number unity Richtmyer-Meshkov flow.  An approximate solution
is derived which is valid throughout the region occupied by the dense
fluid, from bubble to spike, and which captures many of the features
of the late-time flow.  We then compare our theory to the results of a
fluid dynamical simulation.

\section{Previous studies}

Since the pioneering study of Layzer\cite{1955ApJ...122....1L}, the
growth of Richtmyer-Meshkov instability has commonly been described in
terms of the growth of `bubbles' and `spikes', i.e.\@ intrusions of
low-density material into high-density material, and vice versa.  The
width of the mixing region tends to be dominated by the dynamics of
the spikes, while the mass content of the mixing region is dominated
by the bubbles.

Zhang\cite{PhysRevLett.81.3391} generalized the work of Layzer to
treat the asymptotic growth of RM and RT spikes as well as bubbles,
and Sohn\cite{sohn2003simple} extended this analysis for general
Atwood numbers.  The Layzer-type approach followed in both these
papers assumes that the velocity potentials are separable in an
Eulerian frame, and finds solutions for the bubble and spike growth by
matching solutions close to their apexes.
Goncharov\cite{goncharov2002analytical} also extended the formalism to
treat Rayleigh-Taylor growth at arbitrary Atwood number, based on the
propagation of the RT bubble.

Abarzhi et al.\@\cite{abarzhi2003dynamics} use a multiple harmonic
potential to improve the comparison of bubble shapes between theory
and simulation for Rayleigh-Taylor instability, and to avoid the
appearance of a mass flux at negative infinity.
Sohn\cite{sohn2012asymptotic} attempted to reconcile the differences
between the multiple harmonic approach of Abarzhi et al.\@ and the
work of Layzer \& Goncharov.

Mikaelian\cite{mikaelian2005richtmyer} discussed the asymptotic form of Richtmyer-Meshkov
instability when the induced velocity perturbations dominate the
compressed initial amplitude.  This
analysis treats a full spectrum of surface modes, but is strictly
valid only while the surface perturbations have linear amplitude.
This complements the present analysis, which concentrates on a single
mode but is valid in the late time limit of strongly nonlinear
amplitudes.

Duchemin et al.\@\cite{duchemin2005asymptotic} studied the relaxation
to late time asymptotic development of Rayleigh-Taylor spikes at
Atwood number unity.  In the present paper, we apply related complex
variable techniques to study the asymptotic behaviour of
Richtmyer-Meshkov instability for the full domain.

The growth of mass with time has recently been discussed by Cherne et
al.\@\cite{cherne2015shock} using molecular dynamics and continuum
flow calculations, with analysis based on earlier theoretical work by
Layzer \& Mikaelian
\cite{1955ApJ...122....1L,mikaelian1998analytic,mikaelian2003explicit,mikaelian2010analytic},
and the model developed by Buttler et al.\@\cite{buttler2012}, on the
basis of their experimental results.  They find that the mass of
material passing into the spike is controlled by the density of the
material and the area of the base of the spike, which can both be
assumed constant, and that the velocity of the material entering the
spike varies as
\begin{equation}
v = {v_0\over 1+t'/\tau},\label{e:vt}
\end{equation}
where $t'$ is the time when the material leaves the surface.  The
value of $v_0$ can be estimated from Richtmyer's formula for RM
surface growth, $v_0 \simeq ak\delta v$ for feature of Atwood number
$\sim 1$, while the decay time $\tau \simeq \lambda/3\pi v_0$, for a
feature of wavelength $\lambda = 2\pi/k$.  The precise relationship
between $\tau$ and $v_0$ and the initial surface properties will vary
as a result of issues such as the shape of the
bubble\cite{cherne2015shock}: however in the present paper, we are
concerned primarily with the functional form of equation~(\ref{e:vt}).

Durand \& Soulard\cite{durand2015mass} have also performed molecular
dynamic simulations of the growth of a single-mode perturbation.
These calculations were performed in a broader domain than that used
by Cherne et al., which allows a more realistic insight into the
particle break up process to be obtained.  However, the limits on
domain size obtainable by atomistic calculation mean that breakup due
to surface tension will be far more rapid than at the larger scales
typical of, e.g., the experiments of Buttler et
al.\@\cite{buttler2012} Durand \& Soulard develop a numerical model
for the mass-velocity distribution within the jet assuming kinematic
expansion from an initial source defined at the edge of the
fragmentation region, which agrees well with their numerical data.

\section{Kinematic analysis of the spike/jet}

As an initial approximation, we will assume that once the material
enters the spike, it remains at constant density and has constant
velocity, then we can derive several properties of the spikes by
combining the formalism of Cherne et al.\@\cite{cherne2015shock} with
the requirement of mass conservation.  The material at a distance $x$
from the free surface at time $t$ will have been launched at a time
$t'$ given by the implicit equation
\begin{equation}
t' = t -{x\over v},
\end{equation}
from which $v$ may be eliminated using equation~(\ref{e:vt}) to yield
\begin{equation}
t' = {t -x/v_0\over 1+x/v_0 \tau}, \label{e:spikekinematics}
\end{equation}
which can also be written in the more symmetric form
\begin{equation}
1+t'/\tau = {1+t/\tau\over 1+x/v_0 \tau}.
\end{equation}
Alternatively, eliminating $t'$, we find that the velocity of material
within the spike at time $t$ can be expressed as
\begin{equation}
v = {v_0+x/\tau\over 1+t/\tau},\label{e:velpos}
\end{equation}
which shows the linear relationship between position and velocity
expected for kinematic flows from instantaneous sources.  Hence while
this model for ejecta production has a continuous source, the form of
the equations is the same as for an instantaneous source, but with a
virtual time offset $-\tau$, and virtual origin below the surface of
the material, at $x = -v_0\tau\simeq -\lambda/3\pi$.

The mass in the spike between $x$ and $x+{\rm d}x$ at time $t$ is
\begin{equation}
\rho A {\rm\,d}x = -\rho A_0 v {\rm\,d}t',
\end{equation}
where $A$ is the area of the spike material at $x$, $A_0$ is the area
at its base, which we assume is constant as a result of the asymptotic
convergence of the bubble shape, and $\rho$ is the density of the
ejecta material, also assumed constant.  Substituting for ${\rm
  d}x/{\rm d}t'$ at constant $t$ from
equation~(\ref{e:spikekinematics}), we find that -- for this
particular source law, and neglecting break-up processes -- the area
of the spike at a position $x$ relative to the surface is {\it
  independent of time}
\begin{equation}
A = {A_0\over 1+x/v_0\tau},
\end{equation}
for positions $x<v_0t$ which the ejected material can reach in time
$t$.  The total spike volume is logarithmically divergent as
$t\to\infty$, as would be expected from consistency with the rate
that mass is added at the base of the spike.

For velocities above the value given by equation~(\ref{e:vt}), the
cumulative mass as a function of velocity at time $t$ is
\begin{equation}
  M(v,t) = \rho A_0 v_0 \tau\ln\left[\max\left({v_0\over v},1+
      {t\over\tau}\right)\right].\label{e:massdist}
\end{equation}
So in this simple model, the cumulative mass above the free surface
velocity $v=0$ diverges logarithmically at late time, but converges
for any finite $v$.  While the term multiplying the logarithm is
controlled by the wavelength of the perturbation, the logarithmic term
is such that the quantity of ejecta at a specific velocity may be more
closely controlled by the amplitude of the perturbation.

\section{Self-similar solution}

Having seen that the solution appears to relax towards a globally
self-similar state at late time, we will now attempt to find a
solution for this asymptotic flow, which will be valid for the full
domain in the limit of Atwood number one.  The similarity solution
will be found in an accelerating frame, the speed of which may be
deduced by considering the motion of the point of separation at the
tip of the bubble, which must be at a constant position in the
self-similar frame.  We will look for solutions in which the interface
tends to a constant shape at late time, as suggested by the analysis
above, with the flow velocities reducing asymptotically to zero.

For a two-dimensional flow, the equations of incompressible flow are
\begin{eqnarray}
\nabla\cdot {\bf u} &=& 0\\
{\partial{\bf u}\over\partial t} + {\bf u}\cdot\nabla {\bf u} &=& - {1\over\rho} \nabla P,
\end{eqnarray}
where the material has density $\rho$, assumed constant, velocity
${\bf u}$ and pressure $P$.  We transform the flow variables into the
similarity frame using the relations
\begin{eqnarray}
  x &\to& {\lambda\over 2\pi}\left[x' - \alpha \ln(t/\tau)\right] \\
  y &\to& {\lambda\over 2\pi} y' \\
  u_x &\to& {\lambda\over 2\pi t}(v_x - \alpha)\\
  u_y &\to& {\lambda\over 2\pi t}v_y, \\
   P/\rho &\to& \left(\lambda\over 2\pi t\right)^2 p
\end{eqnarray}
where $\alpha$ is a dimensionless parameter.  We are interested in the
self-similar asymptotic solution, so time derivatives in the new frame
of reference will be taken to be zero.  This is the singular $t^{-2}$
case of the self-similar variable-acceleration Rayleigh-Taylor flow
problem discussed by Llor \cite{llor2003bulk}.

Using these substitutions, we find
\begin{eqnarray}
{\partial v_x\over\partial x'}+{\partial v_y\over\partial y'} &=& 0\\
v_x{\partial v_x\over\partial x'}+v_y{\partial v_x\over\partial y'} 
-v_x &=& -{\partial p\over\partial x'} - \alpha\\
v_x{\partial v_y\over\partial x'}+v_y{\partial v_y\over\partial y'} 
-v_y &=& -{\partial p\over\partial y'},
\end{eqnarray}
where for the remainder of the section will suppress the prime on the
accelerated frame variables $(x',y')$.  For boundary conditions we
will assume that $v_y=0$ by symmetry at $y=n\pi$ for integer $n$.  The
apex of the bubble will be at $y=0$, but the $x$ position is left as
arbitrary for the time being.  $v_x=\alpha$, $v_y=0$ is an obvious
trivial solution to this system, in which there is no bubble.

Writing $v_x=\partial\psi/\partial y$, $v_y=-\partial\psi/\partial x$
since $\nabla\cdot {\bf v}=0$ allows Kelvin's theorem to be derived in the
2D accelerating frame, i.e.\@ if we write
\begin{equation}
\omega ={\partial v_x\over\partial y}-{\partial v_y\over\partial x}
\end{equation}
we find
\begin{equation}
\omega - {\bf v}\cdot\nabla\omega = 0,
\end{equation}
which for the boundary conditions of the self-similar problem implies
$w=0$ throughout the domain.  So we then can then write
\begin{eqnarray}
{\partial\over\partial x}
\left[{1\over 2}\left(v_x^2+v_y^2\right)+p\right] &=& v_x-\alpha\label{e:ssberna}\\
{\partial\over\partial y}
\left[{1\over 2}\left(v_x^2+v_y^2\right)+p\right] &=& v_y.\label{e:ssbernb}
\end{eqnarray}
As the flow is irrotational, we can define a velocity potential $\varphi$
such that ${\bf v}=\nabla\varphi$, with the result that
equations~(\ref{e:ssberna}) and~(\ref{e:ssbernb}) may be integrated to
find
\begin{equation}
{1\over 2}\left(\varphi_{,x}^2 + \varphi_{,y}^2\right)+p = \varphi - \alpha x + 
{\it const}.
\label{e:potl}
\end{equation}
This is Bernoulli's equation, modified as result of the accelerating
frame, and allows the pressure distribution to be derived for an
arbitary velocity field.

The remaining condition to be satisfied is the divergence constraint
$\nabla^2\varphi=0$.  The velocity field may be defined by solving this
Laplace equation for $\varphi$ subject to the boundary conditions.  We
are interested in solutions for $\varphi$ which are periodic analytic
functions of $z = x+iy$ in strips of this plane, apart from a region
inside the boundary $p=0$ where poles and branch cuts are allowable,
and with asymptotic behaviour $v_x = \alpha$ as $x\to -\infty$ and
$v_x\propto x$ in the jet regions as $x\to+\infty$.

\begin{figure}
\begin{center}
\includegraphics[angle=270,width=\figwidth]{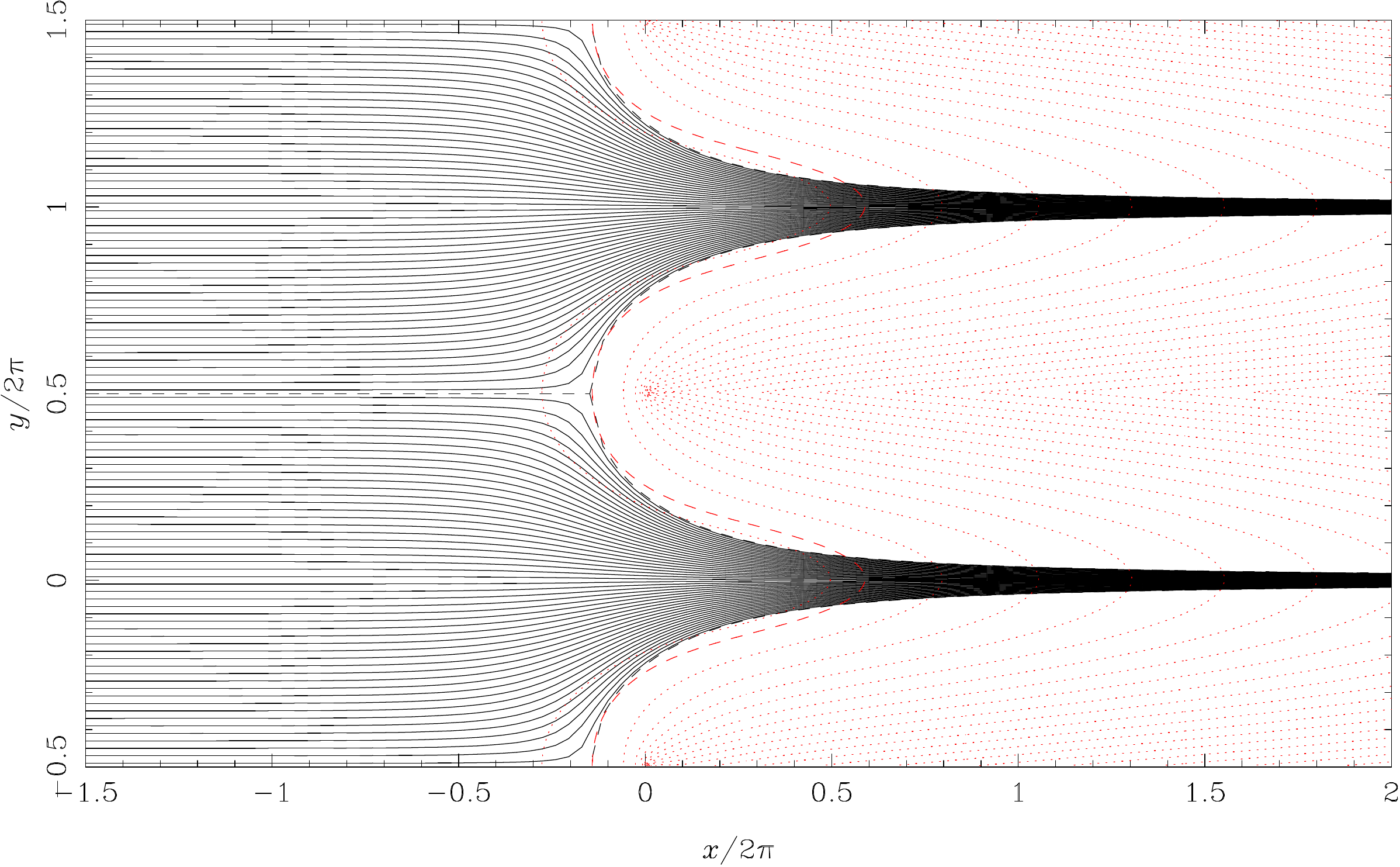}
\end{center}
\caption{Solid black curves show numerical solutions of
  equation~(\ref{e:dilog_ode}) for $\alpha=0.5$, $\beta=1$.  The
  separatrix is shown dashed.  The red dotted curves show contours of
  the pressure, with the red dashed curve being the one at $p=0$ which
  passes through the apex of the bubble.}
\label{f:dilog_1.0}
\end{figure}

Possible solutions of this equation must satisfy the boundary
conditions at $x\to\pm\infty$, periodicity, and the requirement that
the line $p=0$ must be a streamline.  If we initially neglect the last
of these requirements, a fairly broad class of solutions is possible,
functions with branch cuts and poles placed periodically and
symmetrically within the bubbles.  We will study the results of using
the simplest of these
\begin{equation}
v_x = \alpha+\beta\ln(1+\exp x), \label{e:simplevx}
\end{equation}
where $\beta$ is an arbitrary dimensionless constant, which has been
chosen to tend to the usual exponential decay of perturbation velocity
as $x\to-\infty$ but increase linearly as $x\to\infty$.  This form can
be related to the conformal transformation applied by Duchemin et
al.\@ in their study of the asymptotic spike dynamics in Atwood number
unity Rayleigh-Taylor instability.\cite{duchemin2005asymptotic}

Equation~(\ref{e:simplevx}) can be integrated in $x$ and analytically
continued to give $\varphi = \alpha z-\beta{\rm Li}_2(-\exp z)$, where
${\rm Li}_2$ is the dilogarithm
\begin{equation}
{\rm Li}_2(z) = \sum_{k=1}^\infty {z^k\over k^2}
= -\int_0^z {\ln(1-u)\over u}{\rm\,d}u.
\end{equation}
The power series expansion means that the potential can be considered
as a particular form of the multiple harmonic
expansion\cite{abarzhi2003dynamics,sohn2012asymptotic}, chosen to have
acceptable behaviour at $x\to\infty$ to allow for the behaviour of the
spikes.  However, the appearance of a branch cut within the excluded
flow gives the solution a rather distinct character.

Differentiating this analytic function again gives the velocity
distribution
\begin{eqnarray}
  v_x &=& \alpha+{\beta\over 2}
  \ln\left[1+2\cos(y) \exp(x) + \exp(2x)\right] \label{e:vxdilog}\\
  v_y &=& -\beta\arctan\left[\sin y\over \cos y + \exp(-x)\right].
 \label{e:vydilog}
\end{eqnarray}
We can choose that the branch cut for $x>0$, lies on the lines $y =
(2n+1)\pi$, where $v_y$ is no longer continuous, so that this branch
cut resides within the void region separating the jets.  Streamlines
within the void region will not be continuous with $x\to-\infty$.

An ordinary differential equation for the streamlines may be derived
by dividing equation~(\ref{e:vydilog}) by equation~(\ref{e:vxdilog})
\begin{equation}
{{\rm d}y\over {\rm d}x} = 
{-\arctan\left[\sin y/\left( \cos y + \exp(-x)\right)\right]
\over (\alpha/\beta)+{1\over 2} \ln\left[1+2\cos(y) \exp(x) + \exp(2x)\right]},\label{e:dilog_ode}
\end{equation}
which demonstrates that the streamlines are determined by
$\alpha/\beta$.  Some numerical solutions of this equation are shown
in Figure~\ref{f:dilog_1.0}.  We have confirmed that these solutions
are coincident with the contours of ${\rm I}m(\varphi)$.  We evaluate the
dilogarithm function using the methods described by Vollinger \&
Weinzierl.\cite{dilogeval2005}

The values of the dimensionless parameters $\alpha$ and $\beta$ may be
constrained by considering the boundary conditions at the apex of the
bubble, where $y = \pm\pi$, and $v_x=p=0$.  The apex is at $x_a$,
which must satisfy $v_x(x_a+i\pi) = 0$, i.e.\@
\begin{equation}
1-\exp x_a = \exp -\alpha/\beta,
\end{equation}
and
\begin{equation}
v_y = -\beta\arctan \left[1-\exp\left(-\alpha/\beta\right)\right].
\end{equation}

The boundary condition $p=0$ must be satisfied along the full surface
of the jet.  It would be possible to constrain the lowest harmonic
solution further by solving for the curvature at the head of the
bubble, which could be done by using equations~(\ref{e:ssberna})
and~(\ref{e:ssbernb}) to derive an expression for $v\cdot\nabla p$,
which must be zero close to the apex.  However, given the global
nature of the solution, there is a trade-off between the accuracy of
the approximation at the head of the bubble and at larger distances.
We will illustrate the results for values chosen to give acceptable
accuracy overall.

In Figure~\ref{f:dilog_1.0}, we include pressure contours derived
using equation~(\ref{e:potl}).  For the values which we have chosen,
the pressure contour through the apex lies close to the surface
streamline for a significant distance around the bubble.  Varying
$\alpha$ and $\beta$ allows the contours at $x\to\infty$ to be made
more closely parallel to the surface, at the cost of reducing the
accuracy of the fit to the bubble.

\section{Comparison with flow simulation}

We base our calculations on the shock conditions used by Cherne et
al.\cite{cherne2015shock} They model a shocked copper surface,
assuming that the copper is a perfect fluid with $\gamma=3.0$ and
density $8.93{\rm\,g\,cm^{-3}}$, with a light fluid ahead of it with
$\gamma=5/3$ and density $1.22\times10^{-3}{\rm\,g\,cm^{-3}}$.  The
impinging shock is taken to have Mach number $2.5$.  To obtain a
post-shock pressure consistent with their results, the ambient
pressure is assumed to be $10^5{\rm\,Pa}$, while the shocked gas
pressure is $8.875\times 10^5{\rm\,Pa}$.  We have confirmed that this
gives the shocked density and release density given in section IIIB of
this paper.  We offset the frame of motion of our calculations by
$403.6{\rm\,cm\,s^{-1}}$ so that an unperturbed shocked interface
would be at rest.  The timescales reported are scaled to a
characteristic growth time, which we will denote as $\tau^\star$ to
distinguish it from the self-similar scaling timescale which is
expected to be similar, but may not be identical.  This is defined as
$\tau^\star = \lambda/(3\pi\dot{\eta}_0^b)$, where $\dot{\eta}_0^b =
F_lF_{nl}^b\vert kh_0\Delta u\vert$, with $F_l = 1-\Delta u/2 u_s$ and
$F_{nl}^b = 1/(1+k h_0/6)$.  $h_0$ is the initial amplitude of the
surface perturbations, $u_s$ is the shock velocity, and $\Delta u$ is
the velocity jump of the interface; $F_l$ allows for the effect of the
compression of the initial amplitude, while $F_{nl}^b$ models the
saturation of the bubble velocity at large initial amplitudes.

These calculation presented here have been made with the
finite-difference fluid dynamics code
TURMOIL\cite{1991PhFl....3.1312Y}.  This code is particularly suited
for this study, as a result of its good performance in the low Mach
number limit, for flows such as single-mode Kelvin-Helmholtz
roll-up\cite{thornber2008improved}.  The TURMOIL code also implements
semi-Lagrangian mesh motion and one-dimensional Lagrangian boundary
regions, which we use to allow the mesh to move with the mean material
velocity, and hence to minimize advective dissipation.

We concentrate on the low initial amplitude case $kh_0 = 1/8$ to
minimize jetting from the apex of the bubble\cite{cherne2015shock},
which can be driven by post-shock reverberations in the dense
material.  As in this previous work, the calculations presented have
been run with a lateral resolution of 256 cells per wavelength, and
with a large domain size in the direction of shock propagation to
avoid boundary effects: this requires particular care for these
calculations, as a result of the high sound speed in the more diffuse
gas.

\begin{figure}
\begin{center}
\includegraphics[width=\figwidth,clip=true,viewport=109 166 505 674]{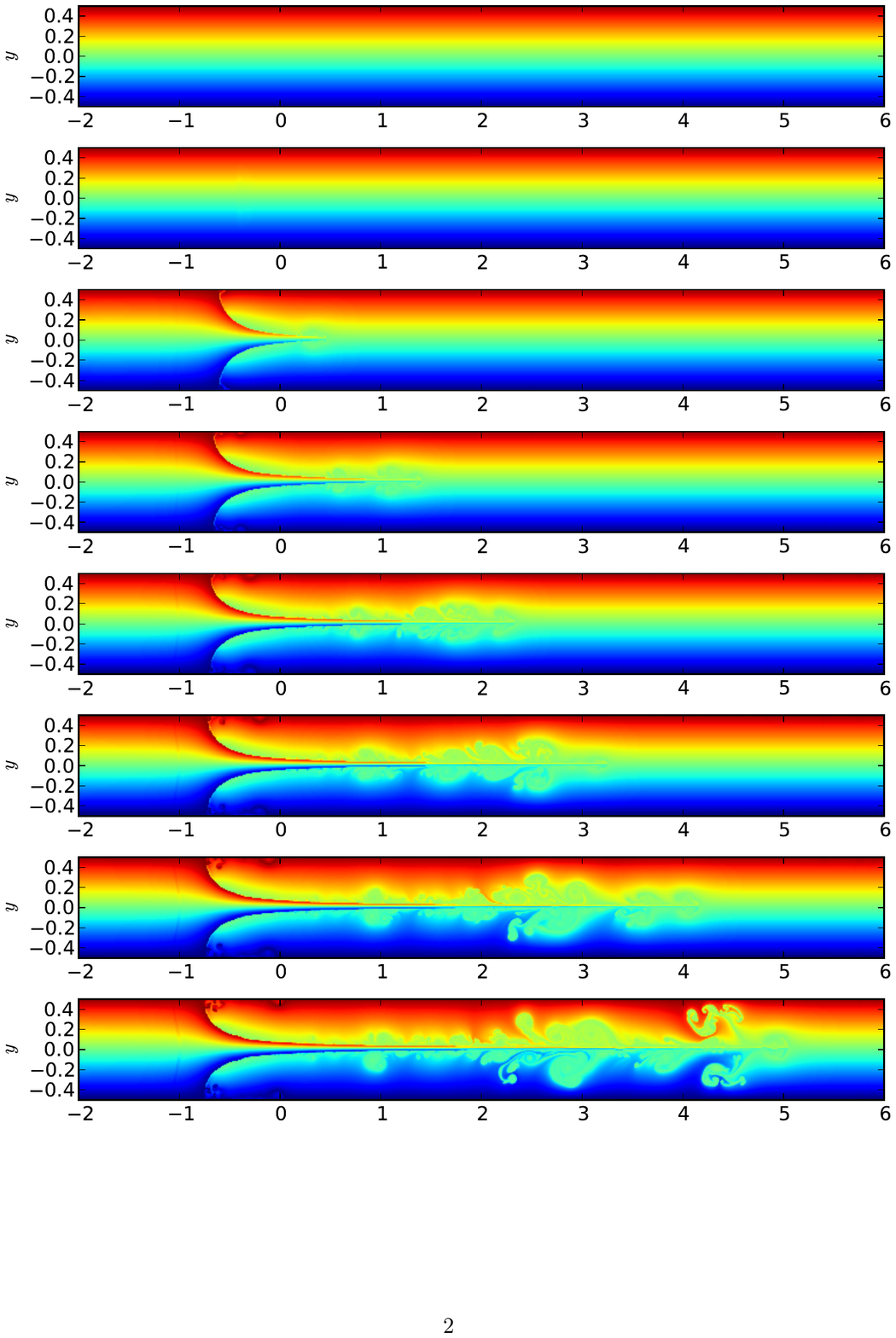}
\end{center}
\caption{Evolution of a tracer variable set to the initial $y$
  position for a $kh_0=1/8$ sinusoidal initial perturbation.  The top
  figure is the initial condition, and subsequent frames are at
  $t/\tau^\star=0,1,2,3,4,5,6$.}
\label{f:simtrace}
\end{figure}

We have also included passively-advected tracer value in the
calculations, with an initial value set by the lateral coordinate
value of the cell.  This allows us to infer the approximate origin of
the material in each cell at later time.  The growth of the
interfacial area with time is determined by a combination of a smooth
straining of the material on the original interface, and the
appearance of new surface material at the apex of the bubble.  In
Figure~\ref{f:simtrace}, we see that the initial surface material has
for the most part been advected into the spike.  As Kelvin's theorem
implies that vorticity is advected with the material, the memory of
the detail of the initial surface profile which determined the
vorticity deposition is also carried into the spike.  Hence, as has
been observed elsewhere,\cite{cherne2015shock} differing surface
profiles will lead to changes in the detailed structure of the spike
tip, but the bubble shape will tend to become universal at late time.

\begin{figure}
\begin{center}
\includegraphics[width=\figwidth,clip=true,viewport=109 166 505 674]{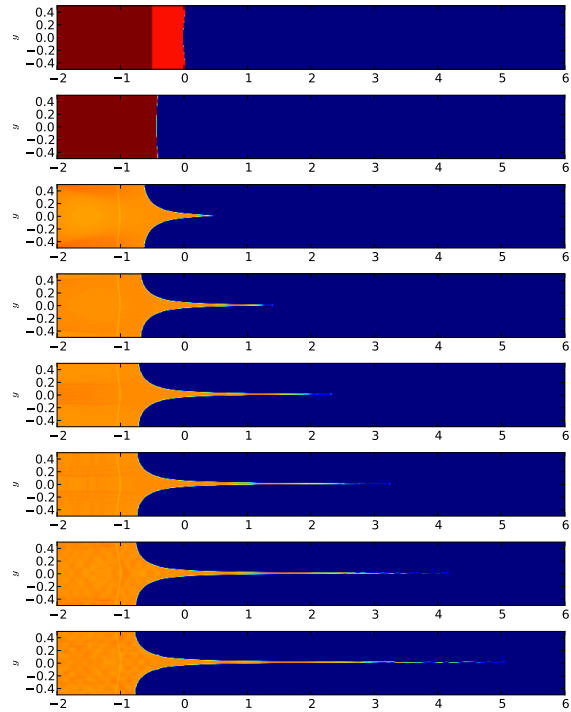}
\end{center}
\caption{Density evolution for a $kh_0=1/8$ sinusoidal initial
  perturbation.  The top figure is the initial condition, and
  subsequent frames are at $t/\tau^\star=0,1,2,3,4,5,6$.  The narrow
  band of low density material at $x=-1$ is caused by a small entropy
  spike deposited at the initial shock position.}
\label{f:simdens}
\end{figure}

\begin{figure}
\begin{center}
\includegraphics[width=0.85\figwidth,clip=true,viewport=136 153 483 694]{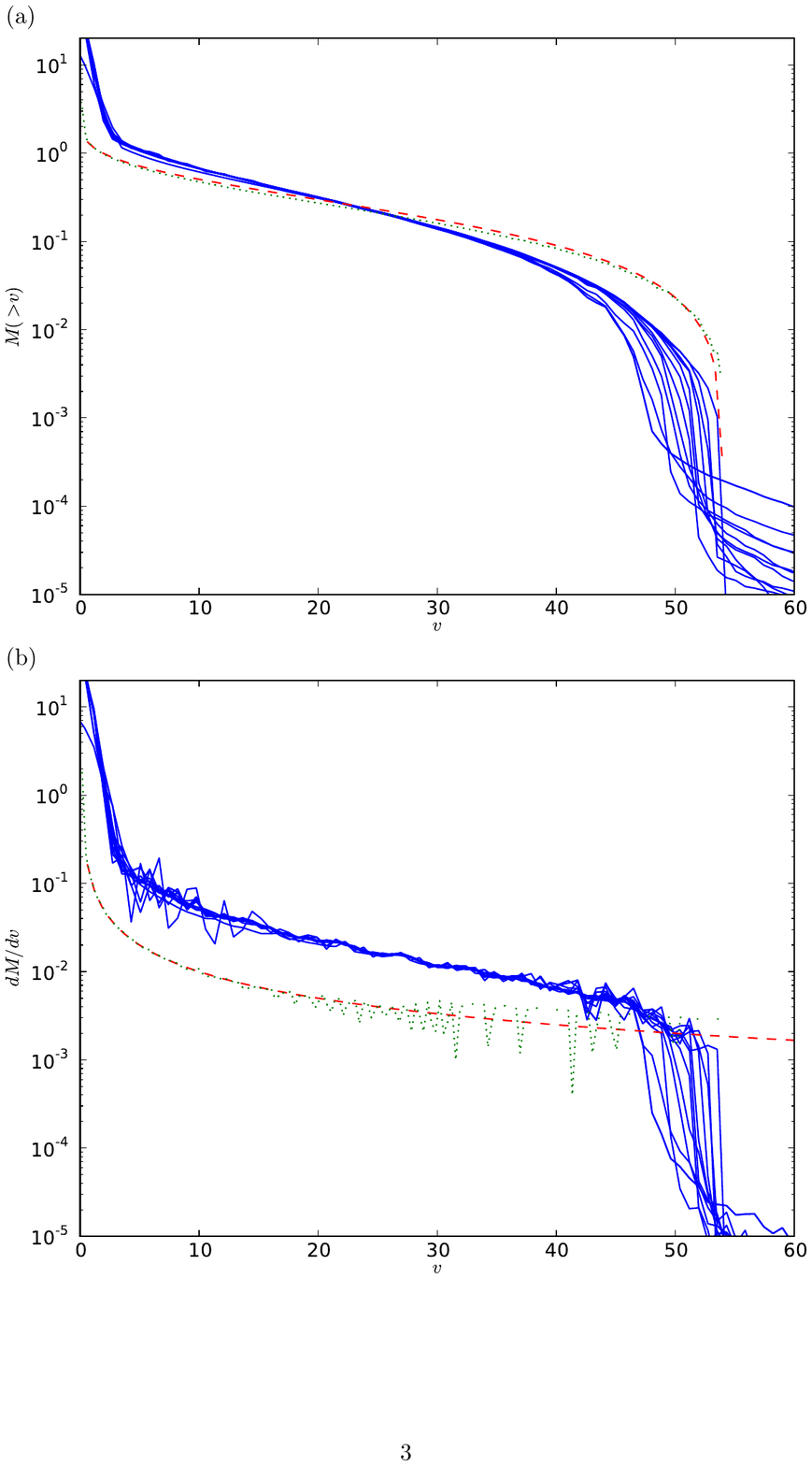}
\end{center}
\caption{Mass-velocity plots from calculations: (a) cumulative, (b)
  resolved.  Blue curves show the calculation results (the knee in
  these curves at high velocity, corresponding to the tip of the
  spike, decelerates somewhat with time).  Dashed red curves show the
  simple model of equation~(\protect\ref{e:massdist}), and dotted
  green curves the similarity solution shown in
  Figure~\protect\ref{f:dilog_1.0} (which is subject to numerical
  noise, as a result of spatial sampling of the solution).  Relative
  normalizations are arbitrary.}
\label{f:cumulative}
\end{figure}

\begin{figure}
\begin{center}
\includegraphics[width=\figwidth]{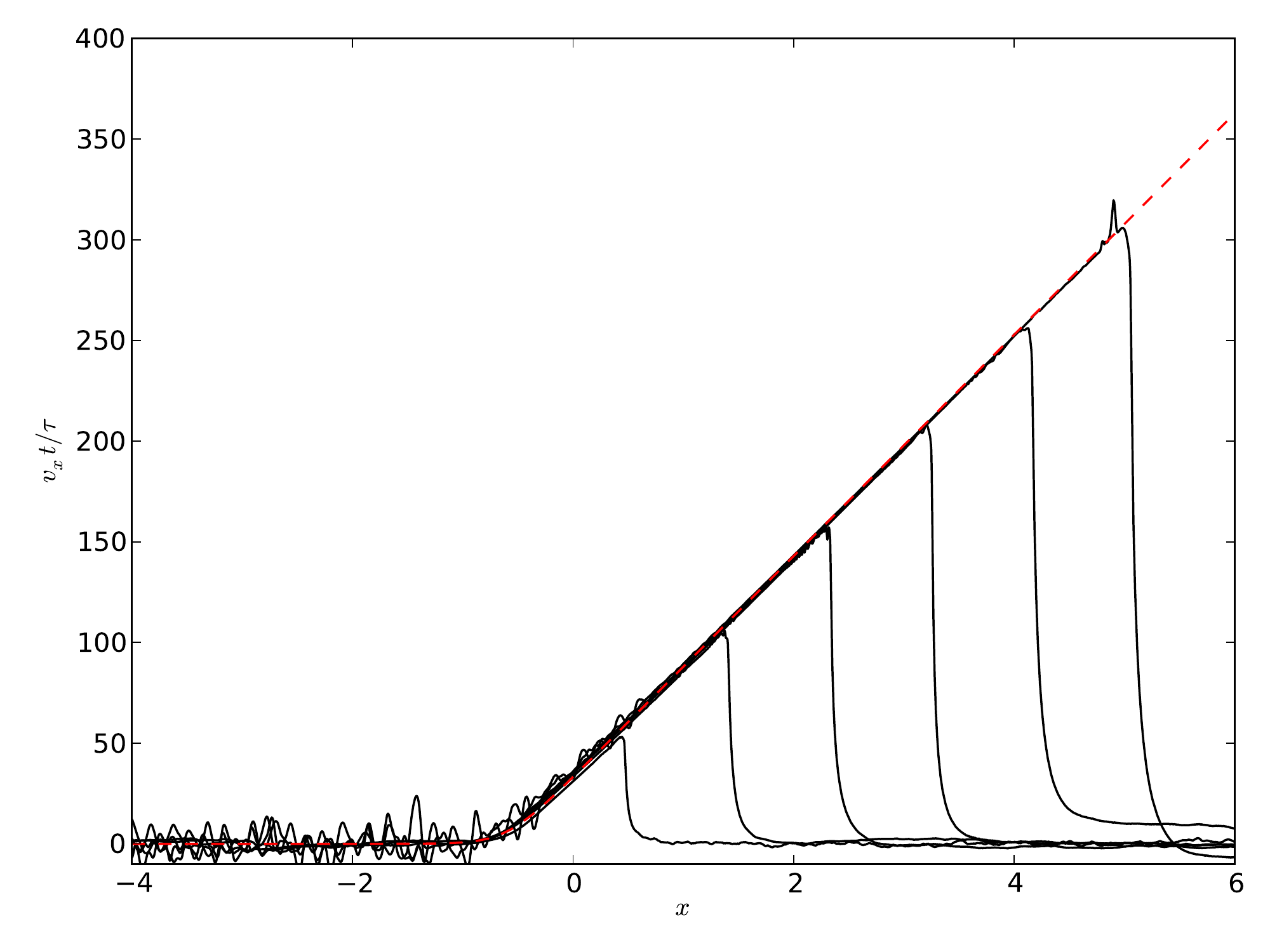}
\end{center}
\caption{Comparison of the flow velocity along the axis of the jet
  scaled into similarity units, $v_x t/\tau^\star$ (solid curves),
  with the axial velocity given by
  equation~(\protect\ref{e:simplevx}), in the frame of the upstream
  material (dashed red curve).  The fluctuations in velocity ahead of
  the spike at the latest times are due to the start of
  vortex shedding from the tip.}
\label{f:asympvel}
\end{figure}

In Figure~\ref{f:simdens} we show the evolution of the material
density with time.  The shape of the bubble is seen to converge to a
form very similar to that of the asymptotic solution in
Figure~\ref{f:dilog_1.0}.  Figure~\ref{f:simtrace} shows the evolution
of the tracer variable set equal to the initial $y$ position of the
material.  As the flow converges, the contours on this plot will tend
to streamlines of the flow.  It can be seen that at late time the
majority of the material on the spike surface has passed through the
stagnation point at the head of the bubble, so information on the
initial surface form encoded by the initial surface vorticity field
will have been advected into the spike.

Vortical structures are seen to develop in the low-density material,
as a result of Kelvin-Helmholtz instability at the surface of the
spike, and at the head of the bubble.  At later times than shown here,
a small spike appears at the head of the bubble, breaking the
self-similar convergence.  This is a familiar phenomenon seen in
previous calculations.\cite{cherne2015shock} It is worth noting,
however, that the decaying stagnation flow at the bubble head is also
likely to be particularly liable to the carbuncle
phenomenon.\cite{quirk1994contribution}

In Figure~\ref{f:cumulative}(a), we compare the cumulative mass
determined from the calculations to the simple model of
equation~(\ref{e:massdist}).  Several timesteps are plotted on this
image, which demonstrates that the mass-velocity distribution has
essentially converged by the first timestep shown, $t/\tau^\star=1$.
The convergence is more rapid than that suggested by
equation~(\ref{e:vt}), presumably because we have included the mass of
all material in the calculation in our evaluation of $M(>v)$, rather
than just the material which has moved beyond the mean free surface.
The model curve is defined by the maximal velocity and overall scale.
The model distribution has a similar form to the calculational data.
The numerical accuracy of the fit is not great; however, the details
of the distribution at the spike tip are expected to be dependent on
the initial surface profile.  The velocity-resolved plot shown in
Figure~\ref{f:cumulative}(b), while more noisy, confirms that
mass-velocity distribution in the similarity solution agrees well with
the analytical form equation~(\ref{e:massdist}).  The comparison with
the results of the numerial calculation suggests the disagreement in
Figure~\ref{f:cumulative}(a) is primarily the result of the structure
at the tip of the spike.

In Figure~\ref{f:asympvel}, we compare the speed along the axis of the
jet scaled into similarity units, $v_x t/\tau^\star$.  This plot
demonstrates the self-similar collapse of the calculation results very
well, with very minor scatter resulting from acoustic modes in the
material upstream of the bubble.  The level of agreement with the
axial velocity given by equation~(\ref{e:simplevx}) here is also very
good, from the bulk material as far as the tip of the spike.

\section{Conclusions}

In the present paper, we have derived a self-similar theory for the
asymptotic behaviour of high density contrast, single-mode
Richtmyer-Meshkov instability.  To match the behaviour of the spikes
at large distances, we have assumed an alternative (but related) form
of velocity potential to that considered in previous studies.  We
propose a simple form for the mass-velocity relationship in the
spikes, equation~(\ref{e:massdist}), which fits the behaviour seen in
calculations to good accuracy, and a more detailed approximate form
for the velocity throughout the dense material,
equations~(\ref{e:vxdilog}) and~(\ref{e:vydilog}).  In contrast to
most previous studies of Richtmyer-Meshkov instability, we model the
late-time asymptotic behaviour of the dense material as a single
domain, rather than treating the bubble or spike in isolation.

As noted above, Abarzhi et al.\@\cite{abarzhi2003dynamics} used a
multiple harmonic expansion to avoid the appearance of a mass flux at
infinity in their model for RT instability.  The potential we assume
includes a mass flux at infinity, but has a similar character to that
considered by Abarzhi et al., in that it has been chosen to match the
asymptotic behaviour of Richtmyer-Meshkov spikes at Atwood number
unity, rather than being exponentially divergent.

The work has focussed upon the growth of two dimensional, single mode
perturbations at high Atwood number.  The development of a general
surface will combine features of the behaviour for regular
perturbations and with those for a localized axisymmetric
perturbations: a localized source at an interface will have a
different scaling behaviour\cite{1991MNRAS.251...93A}.


%
%

%

\begin{acknowledgments}
  The author wishes to thank C.~A. Batha and J.~E. Hammerberg for helpful
  discussions on this work.
  This paper is \copyright{} British Crown Owned Copyright 2016/AWE.
\end{acknowledgments}

\bibliography{rmlate_repr}

\end{document}